\documentclass[aps, prl, twocolumn, notitlepage, superscriptaddress, showkeys, amsfonts, amssymb, amsmath]{revtex4-1}
\usepackage{amsbsy}
\usepackage{amsthm}
\usepackage{float}
\usepackage{graphicx}
\usepackage{latexsym}

\usepackage{color, soul}
\usepackage{ragged2e}

\begin{document}

\title{Agent-based mapping of credit risk for sustainable microfinance}

\author{Joung-Hun Lee}
\affiliation{Faculty of Sciences, Kyushu University, Fukuoka 812-8581, Japan}

\author{Marko Jusup}
\email[Corresponding author: ]{mjusup@gmail.com}
\affiliation{Faculty of Sciences, Kyushu University, Fukuoka 812-8581, Japan}

\author{Boris Podobnik}
\affiliation{Faculty of Civil Engineering, University of Rijeka, 51000 Rijeka, Croatia}
\affiliation{Faculty of Economics, University of Ljubljana, 1000 Ljubljana, Slovenia}
\affiliation{Zagreb School of Economics and Management, 10000 Zagreb, Croatia}

\author{Yoh Iwasa}
\affiliation{Faculty of Sciences, Kyushu University, Fukuoka 812-8581, Japan}

\begin{abstract}
{\bf Inspired by recent ideas on how the analysis of complex financial risks can benefit from analogies with independent research areas, we propose an unorthodox framework for mapping microfinance credit risk---a major obstacle to the sustainability of lenders outreaching to the poor. Specifically, using the elements of network theory, we constructed an agent-based model that obeys the stylised rules of microfinance industry. We found that in a deteriorating economic environment confounded with adverse selection, a form of latent moral hazard may cause a regime shift from a high to a low loan repayment probability. An after-the-fact recovery, when possible, required the economic environment to improve beyond that which led to the shift in the first place. These findings suggest a small set of measurable quantities for mapping microfinance credit risk and, consequently, for balancing the requirements to reasonably price loans and to operate on a fully self-financed basis. We illustrate how the proposed mapping works using a 10-year monthly data set from one of the best-known microfinance representatives, Grameen Bank in Bangladesh. Finally, we discuss an entirely new perspective for managing microfinance credit risk based on enticing spontaneous cooperation by building social capital.}
\end{abstract}


\maketitle

\section*{Introduction}
{\color{black}
Contradictory evidence of the impact of microfinance \cite{Mo99, CBJ01, Kh05, HL11} has simultaneously been the root of high praise and harsh criticism. Proponents argue that the benefit from the access to credit makes microfinance an effective tool for improving the welfare of the poor \cite{KZ10}. Conversely, opponents accuse microfinance institutions (MFIs) of creating even more poverty \cite{Ma03}, while some have gone so far as to claim that MFIs may harbor private groups with vested interest in perpetuating the current situation in many poor regions of the world \cite{ED04}. Both sides, despite ideological differences, have devoted a great deal of debate to the sustainability of microfinance, particularly the reduction of information asymmetries \cite{Mo99, HL07, Co99, McW05} and the non-profit versus for-profit dilemma \cite{Mo99, HL11, HL07, Co99, MS10, McW05}.}

{\color{black}
If an MFI is to operate sustainably, the need to balance reasonable pricing of loans (the social mission) with self-financing through profits (the financial objective) appears to be of utmost importance \cite{Co99, DMM05}. Achieving the desired balance rests on the successful control of credit risk, which is a task complicated by the fact that MFIs often serve a network of borrowers without credit histories. To address the issues involved, we build upon a recent argument that by drawing analogies with independent research areas it is possible to unravel the complexities of financial risks \cite{MLS08, MA10, BRB11, HM11, AKM12}. Specifically, we treated the network of borrowers served by an MFI as a dynamical network, a versatile concept that has proven relevant in studies of many real-world phenomena, including finance \cite{GK07, SFS09, GK10, EGJ14, MPB14, PMC14, PHL14}. The dynamical network was set up to obey the stylized rules of microfinance, where the operations of one of the best established representatives of the industry, Grameen Bank in Bangladesh \cite{Mo99}, constituted a working template.}

\subsection*{Microfinance in a nutshell}
{\color{black}
The distinctive idea of microfinance is outreach to the poor. Every loan manager, accordingly, has the task to recruit customers by visiting their villages and subsequently to organize management units, where each unit covers several villages located near one another. A few units make up a branch under a single manager (i.e. a loan officer in a more traditional context). The importance of such a setup arises from the limited mobility of the potential customers who usually reside away from dense city centers, lack access to credit facilities, and occasionally fall prey to loan sharks.}

{\color{black}
Another key idea behind microfinance is that the intended purpose for the loans is not consumption. Instead, borrowers are supposed to invest into small businesses that can help boost their income. Each borrower is thus expected to justify the loan with a rudimentary business plan. The potential customers typically earn low incomes, less than \$5 US per person or about \$2 to \$30 US per household daily. A representative loan amount is around \$100 US, to be repaid in weekly installments over a period of 52 weeks (one year) with a 20\% nominal interest rate \cite{Mo99}. To overcome the above-mentioned mobility problem, the manager visits each unit once a week to collect the installments.}

{\color{black}
Given that MFIs cannot rely on elaborate credit history calculations, it was necessary to establish microfinance-specific mechanisms for securing the regular payment of installments. Originally, six customers constituted a group, chose a leader from among themselves, and shared liability for loans to the individual members. Because of the shared liability, individual members were subject to peer monitoring and pressure, further reinforced by loan access for other group members being restricted in the case of a default (dynamic incentive). Currently, borrowing implies only individual liability, yet the group structure still exists, presumably to retain some of the benefits of group lending \cite{Mo99, HL07, Co99, St90}. Furthermore, all borrowers belonging to a unit (i.e. about 10 groups) are supposed to attend the weekly meeting with the manager and pay their installments publicly.}

{\color{black}
Other aspects of microfinance intended to encourage borrowers to comply include compulsory savings, dividends, and access to higher loans subject to an exemplary past performance. When borrowers open an account with the bank, they are supposed to save a certain percentage of the loan amount. Although access to these savings is limited, the borrower does obtain certain shareholder rights and is entitled to a dividend. Moreover, borrowers are motivated to stick to the payment schedule because an exemplary performance qualifies them for a progressively higher loan amount in the future. For successful MFIs such as Grameen bank, these mechanisms are sufficient to ensure a payment probability of around 95\% \cite{Mo99, DMM05}.}

\subsection*{Overview of the modeling framework}
{\color{black}
Several assumptions, motivated by Grameen's operations, regulate the network dynamics. First, borrowers are represented as agents (i.e. equal-sized nodes of the network) that can be in an active (paying installments) or an inactive (defaulted) state. The total fraction of active agents, $f$, indicates the overall state of the network. Second, agents can switch from an active to an inactive state with the probability $p_{int}$ to account for the possibility that some borrowers may default due to owning unsuccessful small enterprises. These events are called intrinsic failures. Third, agents are connected (i.e. posses information on each other's state) in the sense of network theory because MFIs entice peer monitoring \cite{Mo99, HL07, Co99, St90} by forming groups of borrowers. Several groups make a unit whose members meet weekly with a loan manager to pay their installments publicly. One manager controls a branch comprised of multiple units from within a geographically limited area. The size of the network is determined by the number of managers employed by a particular MFI. To account for such a structure, we set up a hierarchical network in a way that $n_1 < n_2 < n_3 < n_4$ agents corresponding to a group, a unit, a branch, and the whole MFI, respectively, are connected with the probabilities $q_1 > q_2 > q_3 > q_4$, respectively. Agents connected to each other are referred to as neighbors. Fourth, if a certain critical fraction of neighbors, $t_h$, turns inactive, an agent is tempted---with the small probability $p_{ext}$---to purposely do the same \cite{MPB14, PMC14, Wa02}. Such an extrinsic failure accounts for the latent moral hazard of strategic default when borrowers become wary of the willingness of other borrowers to continue honoring their loan agreements \cite{Mo00, Ma03, CCW07}. The last, fifth assumption of the network dynamics is that the inactive agents in the network, irrespective of whether they failed intrinsically or extrinsically, recover after a time $\tau$ because unsuccessful borrowers may have their payments rescheduled or may be issued new loans to jump-start their businesses. The recovery time has two components, $\tau = \tau_0 + \Delta \tau$, where $\tau_0$ is a certain minimum time needed for loan managers to address the problems of borrowers in default and $\Delta \tau$ is an exponentially distributed random variable \cite{PMC14} with the standard deviation $\sigma$, measuring the diverse abilities of the managers. A complete description of the model can be found in the Methods section, including the derivation of a deterministic analogue. For reference, a summary of mathematical symbols is given in Table~\ref{tab01}. In what follows, our main interest is in applying the introduced dynamical network to identify the effect of the key parameters ($t_h$, $p_{ext}$, $\tau_0$, and $\sigma$) on the fraction of active agents ($f$) given the outside forcing ($p_{int}$).}
\begin{table}[!ht]
\centering
\caption{\bf{List of mathematical symbols in alphabetical order.}}
\renewcommand{\arraystretch}{0.5}
\begin{tabular}{c l}
\hline
\multicolumn{1}{p{1.50cm}}{Symbol} &
\multicolumn{1}{p{6.75cm}}{Description} \\
\hline
$E$                  & prob. of critically inactive neighborhood      \\
$F(l)$               & DFA auto-correlation function                  \\
$F_{ext}(t)$         & number of agents failing extrinsically at $t$  \\
$F_{int}(t)$         & number of agents failing intrinsically at $t$  \\
$F_{tot}(t)$         & total number of agents failing at $t$          \\
$R_{ext}(t)$         & no. of ext. failed agents recovering at $t$    \\
$R_{int}(t)$         & no. of int. failed agents recovering at $t$    \\
$R_{tot}(t)$         & total number of agents recovering at $t$       \\
$c$                  & proportionality constant                       \\
$f$                  & fraction of active agents                      \\
$f^*$                & equilibrium fraction of active agents          \\
$f^*_+$              & high-$f$ equilibrium                           \\
$f^*_-$              & low-$f$ equilibrium                            \\
$l$                  & scale or lag                                   \\
$m_i$                & empirical moments                              \\
$n_{ext}(t)$         & fraction of extrinsically failed agents at $t$ \\
$n_i$                & network size parameters                        \\
$n_{int}(t)$         & fraction of intrinsically failed agents at $t$ \\
$p_{ext}$            & extrinsic failure probability                  \\
$p_{int}$            & intrinsic failure probability                  \\
$\overline{p}_{int}$ & empirical average of $p_{int}$                 \\
$q_i$                & connectedness probabilities                    \\
$s(t)$               & smoothing spline                               \\
$\overline{s}$       & average of $s(t)$                              \\
$t_h$                & fractional threshold                           \\
$\Delta p_{int}(t)$  & zero-mean smooth forcing function              \\
$\Delta\tau$         & random component of $\tau$                     \\
$\alpha$             & scaling exponent of $F(l)$                     \\
$\sigma$             & standard deviation of $\Delta p_{int}(t)$      \\
$\tau$               & recovery time                                  \\
$\overline{\tau}$    & average recovery time                          \\
$\tau_0$             & minimum recovery time                          \\
\hline
\end{tabular}
\label{tab01}
\end{table}

\section*{Results}
\begin{figure}[!ht]
\begin{center}
\scalebox{0.5}{\includegraphics{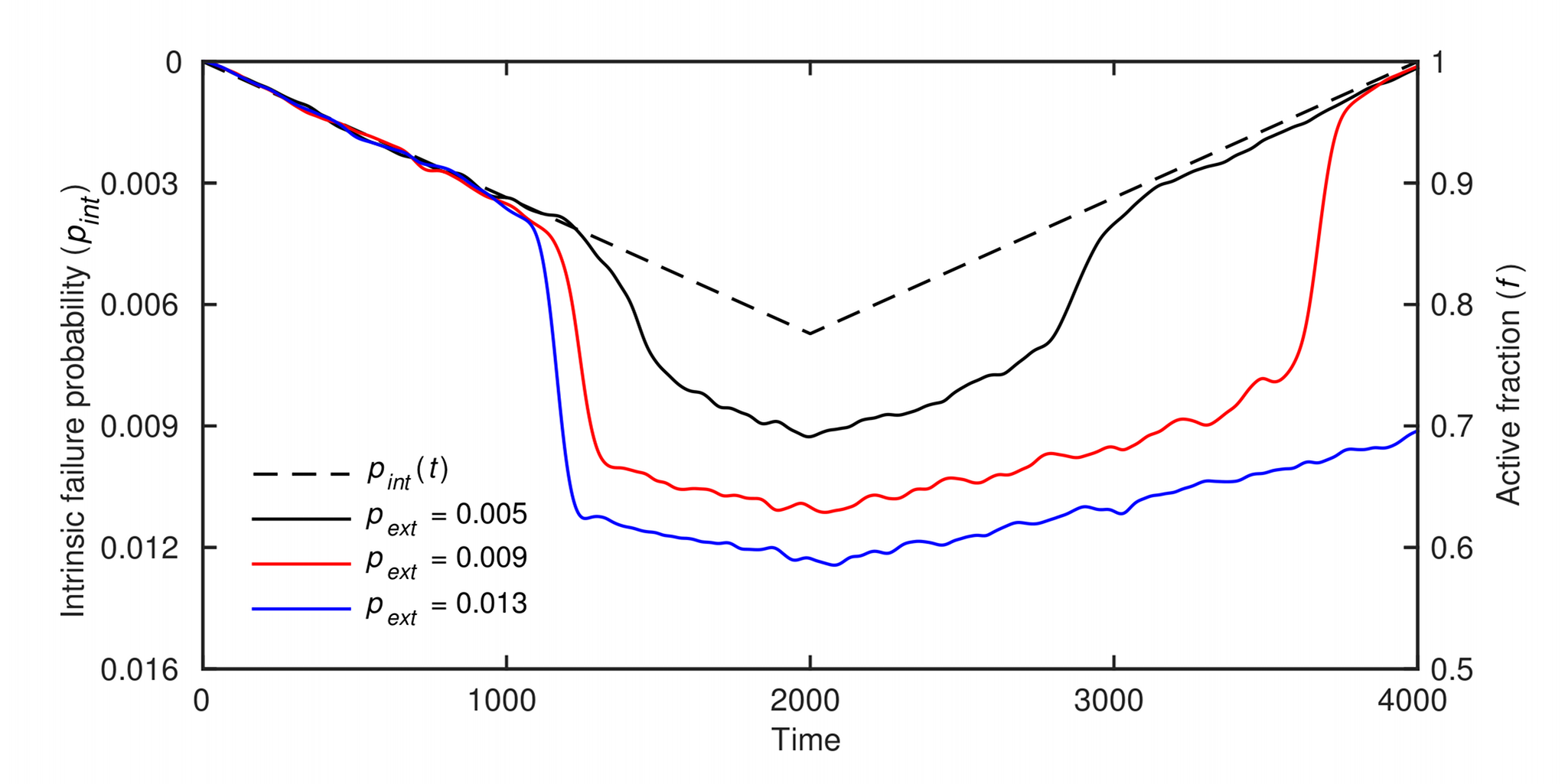}}
\end{center}
\caption{{\bf Time evolution of the dynamical network in a stylized economic downturn.} The probability of intrinsic failure, $p_{int}$, steadily increases (decreases) throughout the first (second) half of the simulation at a rate $3.854 \times 10^{-6}$ per time-step, symbolizing a deteriorating (recovering) economic environment. The network is seen switching between regimes with high and low fractions of active agents, $f$. Depending on the probability of extrinsic failure, $p_{ext}$, the network performance may lag behind that of the economy (red curve) or it may never return to the high-$f$ regime (blue curve). Other parameter values are $t_h=0.2$, $\tau_0=7$, and $\sigma=30$. In all simulations, $n_1=6$, $n_2=60$, $n_3=420$, with the total number of agents set to $n_4 \approx 10^4$ for computational reasons. The connectedness probabilities are $q_1=1$, $q_2=0.7$, and $q_3=0.05$, and $q_4$ is constrained so that the total average number of connections is 100~\cite{Du92}. The scale for $p_{int}$ is reversed and magnified 32 times for easier visual comparison.}
\label{fig01}
\end{figure}

{\color{black}
To sustainably provide loans to borrowers not served by the traditional financial system, MFIs are critically dependent on a high loan payment probability, itself largely driven by the economic environment in which a given MFI operates. We therefore examined the performance of the network of borrowers within a deteriorating economic environment by using the described dynamical framework (Fig.~\ref{fig01}). For fixed values of $t_h$, $p_{ext}$, $\tau_0$, and $\sigma$, the results showed that the performance as measured by the fraction of active agents ($f$) responds linearly to the probability of intrinsic failure ($p_{int}$) up to a certain point. When the value of $p_{int}$ increases sufficiently, the high rate of intrinsic failures drives the fraction of inactive neighbors of some agents close to the critical value, $t_h$, causing extrinsic failures to occur for the first time. The dynamical network thereafter undergoes a regime shift evident from a highly nonlinear decline in $f$ (Fig.~\ref{fig01}). Additionally, the recovery of the network can be delayed, meaning that $p_{int}$ must be lower than the value that caused the downward regime shift before the fraction of active agents undergoes the reverse shift. In some cases, recovery is altogether absent. To understand these patterns better, we proceeded with a systematic analysis of the dynamics.}

\begin{figure}[!ht]
\begin{center}
\scalebox{0.5}{\includegraphics{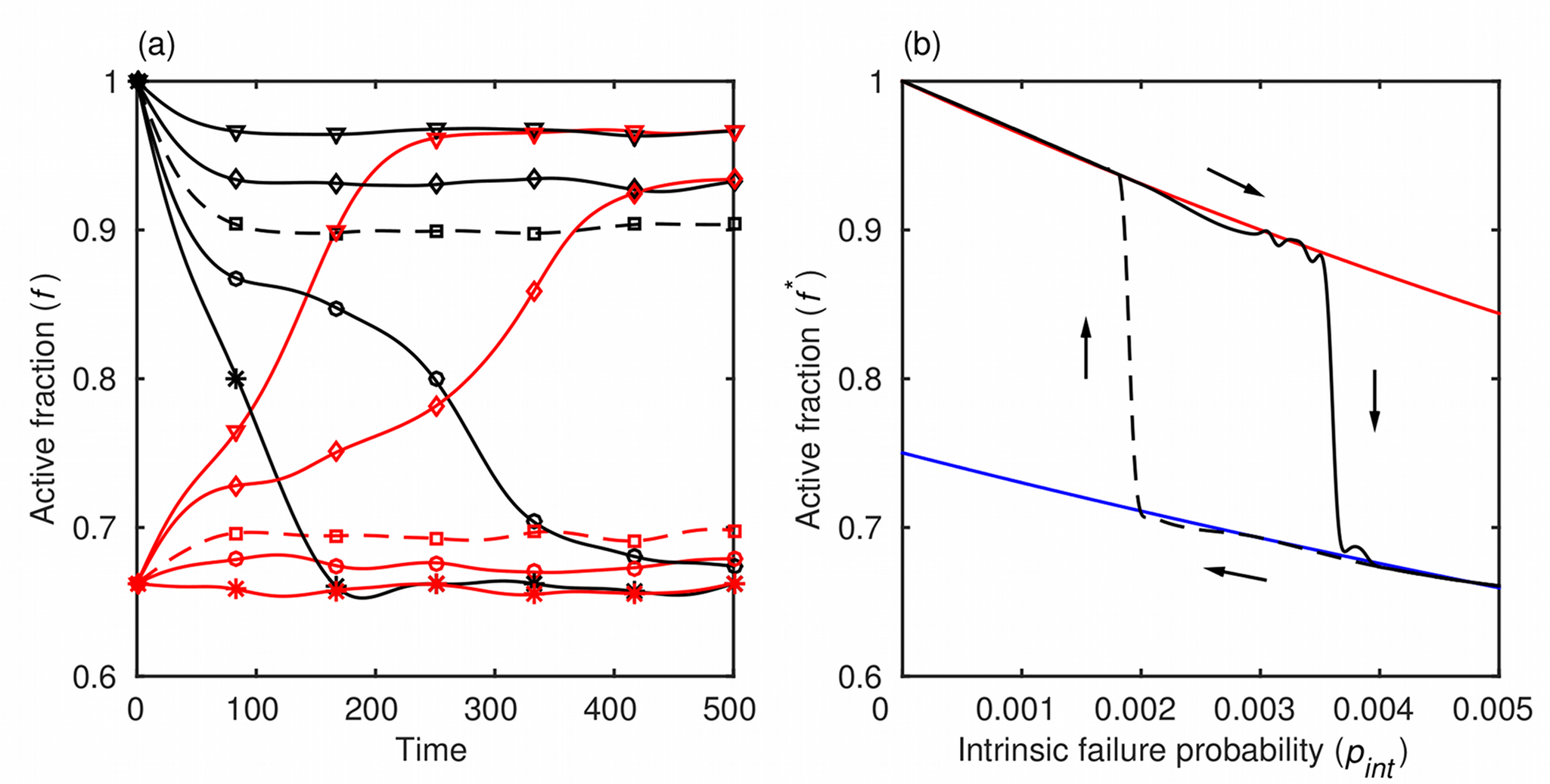}}
\end{center}
\caption{{\bf Two types of equilibria and the regime shift.} (a) The performance of the dynamical network as measured by $f$ converges either to a high-$f$ or a low-$f$ equilibrium, depending on the probability of intrinsic failure, $p_{int}$,  and the initial state. The parameter $p_{int}$  ranges from 0.001 (triangles) to 0.005 (stars) in increments of 0.001. The dependence on the initial state is exemplified by the dashed curves, which converge to different equilibria although both were generated with the same $p_{int}$. (b) The equilibrium states ($f^*$) of the dynamical network form a hysteresis loop. Numerical results (black) compare favorably with the analytically derived high-$f$ and low-$f$ equilibria (red and blue curves, respectively; see Eq.~(\ref{eq01}) and its limits in the text). A major consequence of hysteresis is that the recovery of the network lags that of the economy, because the probability of intrinsic failure, $p_{int}$, causing a shift from a high-$f$ to a low-$f$ regime is larger than the corresponding probability permitting the opposite regime shift. Parameter values are $t_h=0.2$, $p_{ext}=0.009$, $\tau_0=7$, and $\sigma=30$.}
\label{fig02}
\end{figure}

{\color{black}
We began the analysis by observing how the fraction of active agents changes with time when the probability of intrinsic failure, $p_{int}$, is kept constant (Fig.~\ref{fig02}a). Several important aspects of the dynamical network are revealed this way. The performance as measured by the fraction of active agents, $f$, converges to an equilibrium, $f^*$. There are two types of equilibria, one with a high and one with a low fraction of active agents, denoted $f^*_+$ and $f^*_-$, respectively. Convergence to a high-$f$ or a low-$f$ equilibrium depends on the value of $p_{int}$ and, somewhat surprisingly, on the initial state, in which all inactive agents are assumed to have failed intrinsically and have had their recovery time drawn from the exponential distribution with the parameter $\sigma$. The dependence on the initial state is exemplified by the two dashed curves in Fig.~\ref{fig02}a, which, although generated with the same value of $p_{int}$, converge to equilibria of different types. Before turning to the implications of these outcomes, we first explore the origin of the two types of equilibria.}

{\color{black}
In the equilibrium state, by definition, rates at which agents fail and recover are the same. As long as the overall equilibrium fraction of active agents in the network, $f^*$, is sufficiently high, the majority of agents will be aware of a fraction of inactive neighbors below the critical value $t_h$. In this case, all failures are intrinsic and the equilibrium failure rate is simply $f^* p_{int}$. Worsening performance of the network complicates matters because it increases the probability (denoted $E$) of having a critically inactive neighborhood, i.e. the probability that a randomly chosen agent has a fraction of inactive neighbors higher than $t_h$. Therefore, in addition to intrinsic failures, the remaining active agents in the equilibrium state, $f^* (1 - p_{int})$, can fail extrinsically, increasing the total failure rate by $f^* (1 - p_{int}) E p_{ext}$. The recovery rate is determined by the reciprocal of the average time to recovery, $\overline{\tau} = \tau_0 + \sigma$, and the fraction of inactive agents, $1 - f^*$, producing the term $(1 - f^*) / \overline{\tau}$. By balancing failure and recovery rates, and solving for $f^*$, we get
\begin{equation}
 f^* = \frac{1}{1 + (p_{int} + E p_{ext} - E p_{int} p_{ext}) \overline{\tau}}.
\label{eq01}
\end{equation}
From this equation, in the limits $E \to 0$ and $E \to 1$, we obtain high-$f$ ($f^*_+$) and low-$f$ ($f^*_-$) equilibria, respectively. When the conditions $p_{int} \overline{\tau} \ll 1$ and $E p_{ext} \overline{\tau} \ll 1$ are satisfied, a series expansion of Eq.~(\ref{eq01}) yields $f^* \approx 1 - (p_{int} + E p_{ext} - E p_{int} p_{ext}) \overline{\tau}$, which is the result reported by Ref.~\cite{MPB14} after assuming independent intrinsic and extrinsic failures. This mean-field approximation works much better for regular, as opposed to Erd\H{o}s-R\'{e}nyi, networks \cite{MPB14}, and the larger the system the more appropriate it becomes. The other terms in the series (not shown) originate from the fact that intrinsic and extrinsic failures in our model are dependent. The insight gained from Eq.~(\ref{eq01}) and its limits puts us in a better position to understand the implications of the previously described outcomes, but before discussing these implications, we first present an additional visual aid.}

{\color{black}
The visual aid in question (Fig.~\ref{fig02}b) is a plot that shows how the equilibrium state ($f^*$) depends on the probability of intrinsic failure ($p_{int}$). The most prominent feature of the plot is that the equilibrium states of the dynamical network form a hysteresis loop. Namely, the critical value of $p_{int}$ causing the shift from a high-$f$ to a low-$f$ equilibrium is larger than the corresponding critical value at which the opposite shift is possible. An immediate consequence is that the convergence depends on the initial state, as demonstrated in Fig.~\ref{fig02}a, or, more generally, on the path traversed by the fraction of active agents over time. Another consequence is the delayed recovery illustrated in Fig.~\ref{fig01}. In fact, when the parameter values are such that the returning branch of the hysteresis loop (dashed curve in Fig.~\ref{fig02}b) gets pushed beyond the range of possible forcings (i.e. below $p_{int}=0$, which is accomplished, for instance, by increasing $p_{ext}$), a recovery is no longer observed. We next argue that important insights into microfinance credit risk management follow from these results and extensions thereof, shown in the form of the phase diagrams.}

\begin{figure}[!ht]
\begin{center}
\scalebox{0.5}{\includegraphics{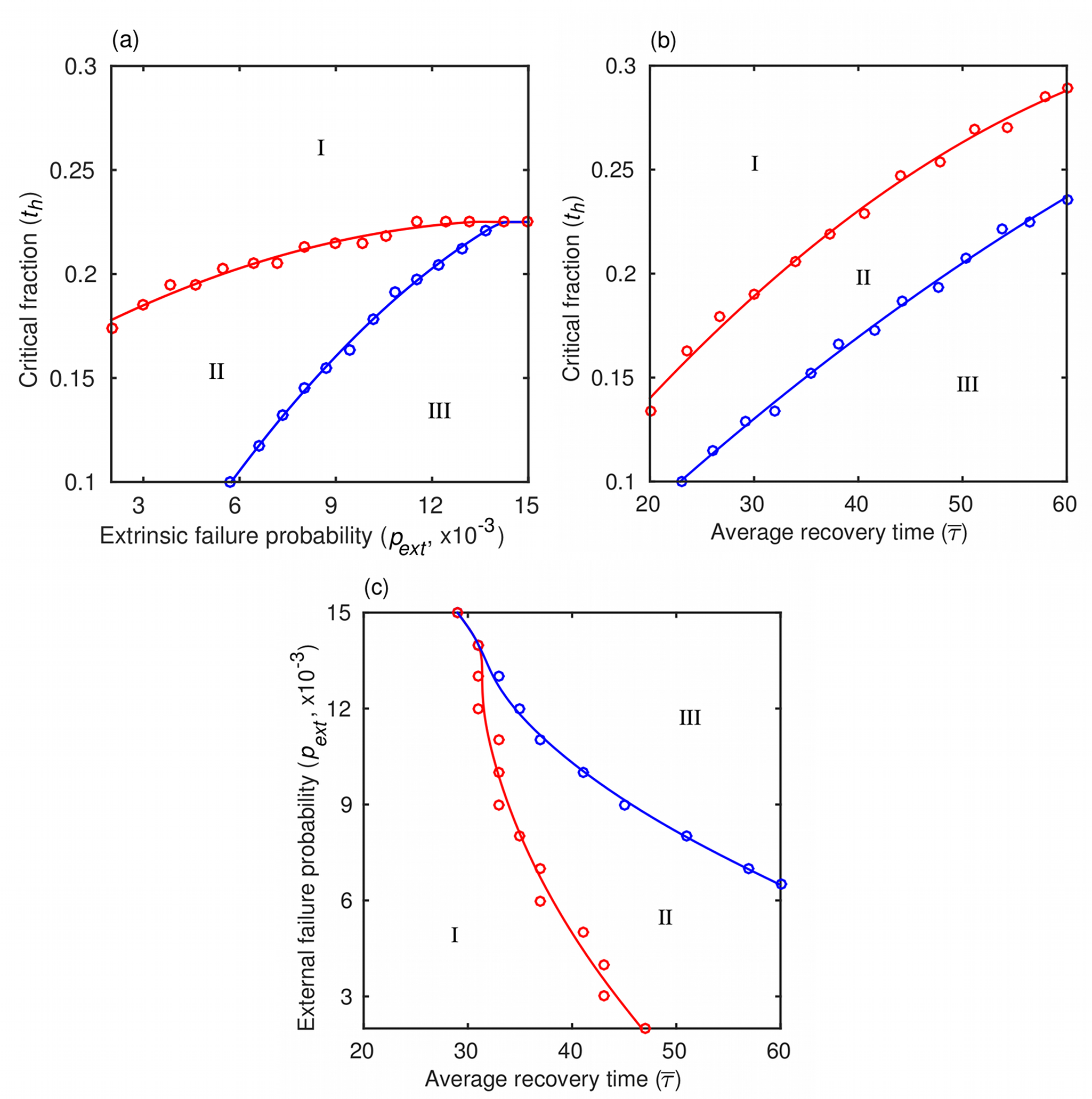}}
\end{center}
\caption{{\bf Risk maps for credit risk evaluation.} We are interested in where the real-world lender's parameters are situated relative to the region of the phase space characterized by high instability. Our model depends on three parameters, but instead of a three-dimensional phase space, we plot risk maps in three separate two-dimensional phase planes. The model exhibits three distinct behaviors depending on the parameter values: a high-$f$ regime (I), a low-$f$ regime with recovery (II), and a low-$f$ regime without recovery (III). (a) Phase diagram in the $t_h$-$p_{ext}$ plane with constant $\overline{\tau} = \tau_0 + \sigma$, where $\tau_0=7$ and $\sigma=30$. (b) Phase diagram in the $t_h$-$\overline{\tau}$ plane with constant $p_{ext}=0.009$. (c) Phase diagram in the $p_{ext}$-$\overline{\tau}$ plane with constant $t_h=0.2$. The probability of intrinsic failure is $p_{int}=0.004$. Shown are the numerical results and curves of best fit.}
\label{fig03}
\end{figure}

{\color{black}
When estimating risk, it is crucial to find out how far the parameters of a given dynamical system are from the regions of the parameter space characterized by high instability. We therefore use phase diagrams to present a comprehensive overview of the dynamics as a function of the parameter values. These diagrams are referred to as risk maps because of the assumed link between the dynamics and credit risk. For a given value of the probability of intrinsic failure ($p_{int}$), which is, as stated before, beyond the control of MFIs, we can produce three two-dimensional risk maps (Fig.~\ref{fig03}), one in the $t_h-p_{ext}$ plane (constant $\overline{\tau} = \tau_0 + \sigma$), another in the $t_h - \overline{\tau}$ plane (constant $p_{ext}$), and the last one in the $p_{ext} - \overline{\tau}$ plane (constant $t_h$). Each of the three risk maps shows three distinct regimes of the network dynamics. Regime I is characterized by a high equilibrium fraction of active agents, $f^*_+$, thus carrying little credit risk for the MFI. In contrast, regime III has a low equilibrium fraction of active agents, $f^*_-$, such that the network can never re-enter regime I even if the probability of intrinsic failure improves to $p_{int} = 0$. Regime III, consequently, implies a high credit risk for an MFI and needs to be avoided at all cost. Intermediate regime II generally leads to the low-$f$ equilibrium, $f^*_-$, from which it is always possible to recover provided the probability $p_{int}$ improves sufficiently. In addition, for small values of $p_{ext}$ (roughly $<0.005$; not shown), some paths satisfy $f^*_- < f < f^*_+$ for an indeterminate period of time because the fraction of failed neighbors ends up distributed around the critical fraction, $t_h$. Mathematically, we have $0 < E < 1$. These paths always permit a recovery and hence qualitatively belong to regime II. In the context of managing microfinance credit risk, regime II implies a necessity for caution and it should prompt actions aimed at restoring better performance. By classifying the network dynamics into three distinct regimes and identifying their respective implications for credit risk management, we have made the connection between the dynamics and risk more apparent, but we have yet to specify the factors controlling the prevalence of each regime.}

{\color{black}
The risk maps (Fig.~\ref{fig03}) reveal the factors that promote the prevalence of each of the network dynamics regimes. Here, it is useful to recall that the smaller the value of $t_h$ and the larger the value of $p_{ext}$, the more an agent depends on its neighbors. When the average time to agent recovery, $\overline{\tau} = \tau_0 + \sigma$, is constant (Fig.~\ref{fig03}a), a high enough value of $t_h$ always keeps the network in regime I. This result is understandable given that the probability of extrinsic failure, $p_{ext}$, plays a role in the dynamics only if the intrinsic failure rate is sufficient by itself to raise the fraction of failed neighbors of an agent to the level of $t_h$. If $t_h$ is high enough, we may never observe such an outcome. Conversely, a decreasing value of $t_h$ generally forces the network to jump from regime I to II, unless $p_{ext}$ is sufficiently high to force a direct transition to regime III. Qualitatively different behavior is observed when $p_{ext}$ is held constant and $\overline{\tau}$ is allowed to change (Fig.~\ref{fig03}b). In this case, for any value of $t_h$, recoveries from failures can be sufficiently slow (i.e. $\overline{\tau}$ sufficiently high) that the network transitions first from regime I to II and then from regime II to III. However, direct transitions from regime I to III are impossible. Finally, when $t_h$ is held constant (Fig.~\ref{fig03}c), an increasing $\overline{\tau}$ first drives the network from regime I to II and then from regime II to III, unless again $p_{ext}$ is sufficiently high for a direct transition to regime III to be possible. In addition to mapping microfinance credit risk in the described manner, we can also consider the question if there are any early-warning signs that could be exploited to mitigate or even avoid the negative impact of a regime shift.}

\begin{figure}[!ht]
\begin{center}
\scalebox{0.5}{\includegraphics{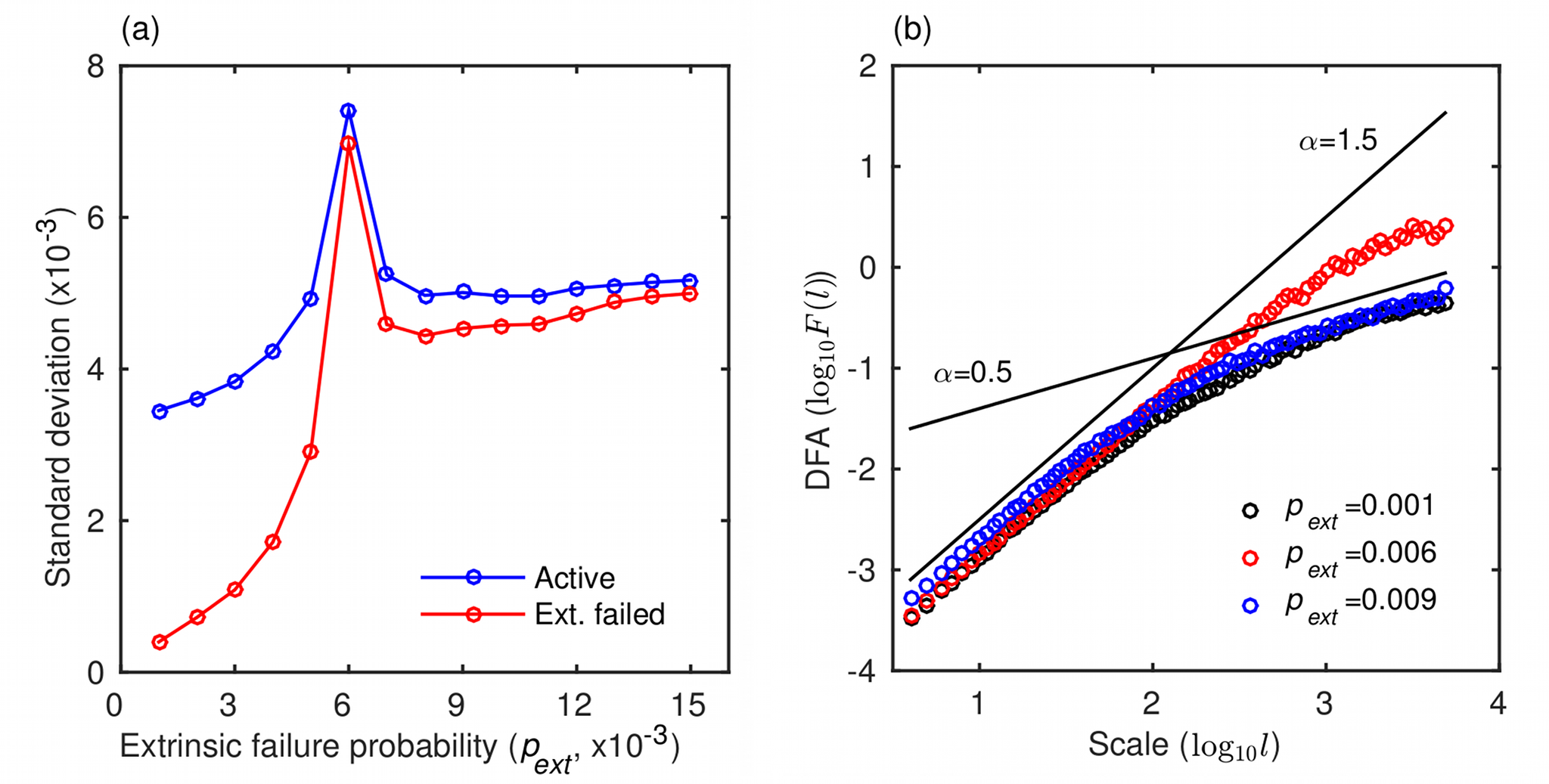}}
\end{center}
\caption{{\bf Early-warning indicators of a regime shift.} (a) Standard deviations of the fractions of active and extrinsically failed agents spike as the dynamical network undergoes a shift from regime I to regime II at $p_{ext}=0.006$ (cf. Fig.~\ref{fig03}a, c). (b) Auto-correlation at the network level due to agent failures exhibits a longer memory close to the regime shift at $p_{ext}=0.006$ (cf. Fig.~\ref{fig03}a, c). In detrended fluctuation analysis (DFA) the strength of the correlation is reflected in the scaling exponent, $\alpha$ (corresponding to the slope in the logarithmic plot). Lines with slopes $\alpha=1.5$ and $\alpha=0.5$ (no correlation) are plotted for easier visual comparison. Parameter values are $t_h=0.2$, $\tau_0=7$, and $\sigma=30$. The probability of intrinsic failure is $p_{int}=0.004$.}
\label{fig04}
\end{figure}

{\color{black}
As a critical point is approached, complex dynamical systems often become less attracted to the equilibrium state, experiencing bigger displacements and slower recoveries from perturbations \cite{Sc10}. This phenomenon---dubbed critical slowing down---should, therefore, reveal itself through a larger standard deviation and longer correlation of the state variables. In Fig.~\ref{fig04}a, we show that a gradual increase in the probability parameter $p_{ext}$ causes the standard deviations of the fractions of active and extrinsically failed agents to spike when the dynamical network approaches a critical point during the shift from regime I to regime II. Because spiking begins noticeably before the regime shift, this critical slowing down may be used as an early-warning indicator of a network breakdown \cite{Sc10}. We also conducted a detrended fluctuations analysis (DFA) \cite{PBH94} to analyze auto-correlation in the model outputs (Fig.~\ref{fig04}b). The DFA auto-correlation function is of the form $F(l) \propto l^{\alpha}$, where $l$ is the scale (lag) and $\alpha$ is an exponent measuring the strength of the correlation. When $\alpha < \frac{1}{2}$ ($\alpha > \frac{1}{2}$) two consecutive displacements of the network state---e.g. of the fraction of active agents $f$---are more likely to be in the opposite (same) direction. The exponent $\alpha = \frac{1}{2}$ indicates no correlation. In our dynamical network, for small $p_{ext}$, we observe (Fig.~\ref{fig04}b) a finite-range auto-correlation of the random-walk type ($\alpha \approx 1.5$), but its strength gradually diminishes at larger scales \cite{PMC14}. However, as the network undergoes the regime shift at $p_{ext}=0.006$, the auto-correlation exhibits considerable strength even at large scales. Once $p_{ext}$ takes the network decisively into regime II, the strength of the auto-correlation is again similar to that observed in the case of small $p_{ext}$. Accordingly, we identified two potential early-warning indicators complementary to the microfinance credit risk mapping, both of which could help mitigate the negative impact of a regime shift.}

\subsection*{Empirical analysis}
\begin{figure}[!ht]
\begin{center}
\scalebox{0.5}{\includegraphics{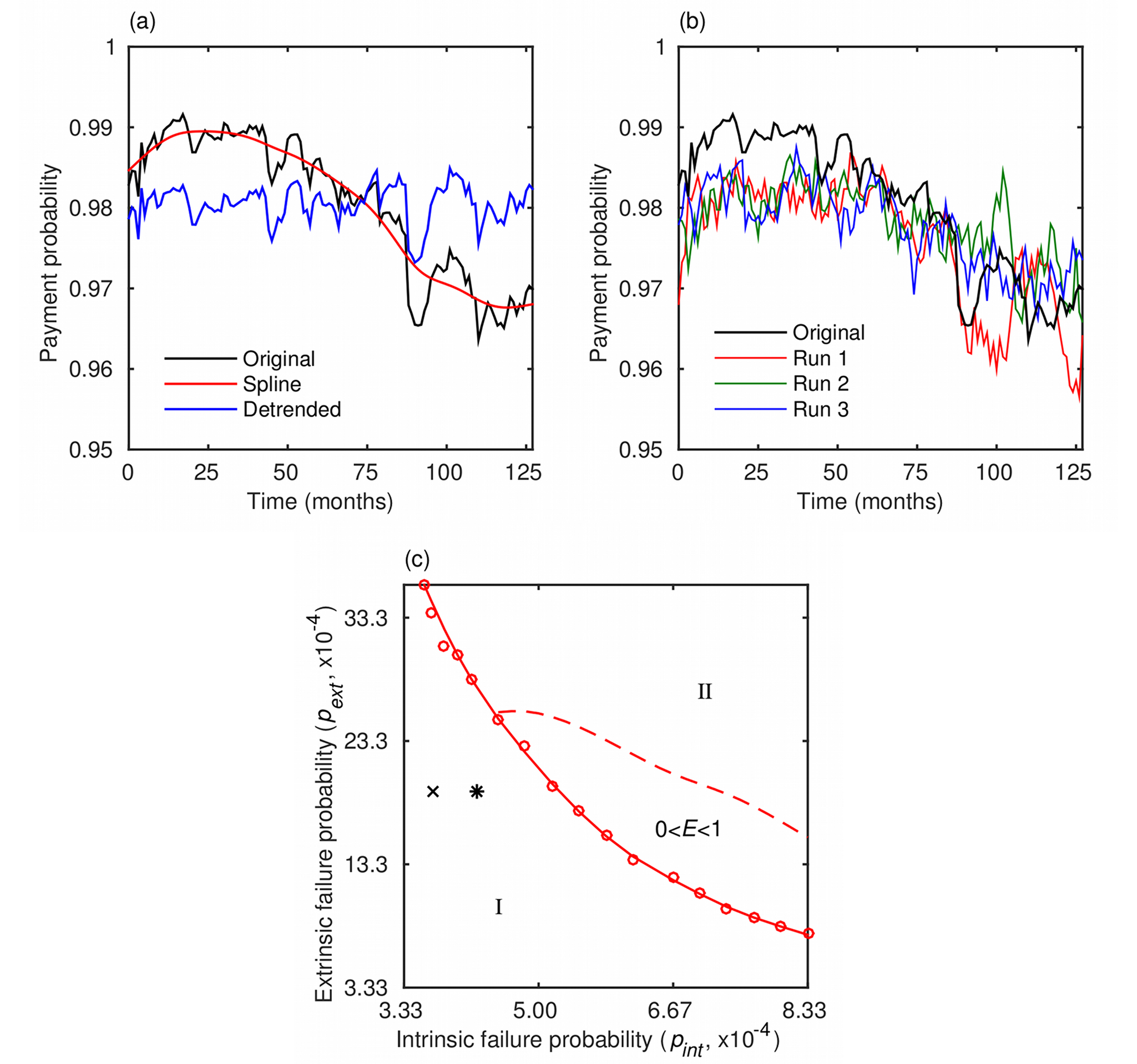}}
\end{center}
\caption{{\bf Data set, comparison with model runs, and the risk map for Grameen bank.} (a) The data set contains monthly information on amounts due that were actually paid to Grameen bank between June 2002 and Jan 2013. A long-term trend observable in the data is extracted using a smoothing spline, $s(t)$. The trend suggests that the model forcing is of the form $p_{int}(t) = \overline{p}_{int} + \Delta p_{int}(t)$, where $\overline{p}_{int}$ is a constant and $\Delta p_{int}(t)$ a zero-mean smooth function of time. (b) Comparison of the data set with typical model runs reveals satisfactory agreement between the two. For the forcing term, we assume $\Delta p_{int}(t) = c [s(t) - \overline{s}]$, where $c=-\frac{1}{8}$ and $\overline{s}$ is the average of $s(t)$. (c) Estimated model parameters place Grameen bank in regime I (x-mark) as could be expected from one of the most successful microfinance representatives. Even in the wake of 2007-2008 financial crisis, Grameen Bank continues to operate in the low risk regime (star) despite the more challenging economic environment.}
\label{fig05}
\end{figure}

{\color{black}
To interpret the real-world microfinance data (Fig.~\ref{fig05}a) in the context of our modeling framework, we performed an empirical analysis by estimating the model parameters from the available data set using the method of moments (see the Methods section). Upon obtaining the parameter estimates, we conducted simulations to visually compare the data set and the model outputs (Fig.~\ref{fig05}b). The results of the empirical analysis indicate a very small probability of intrinsic failure, $\overline{p}_{int}=3.7 \times 10^{-4}$ per day, which is equivalent to saying that each month about 1.1\% of the Grameen's network of active borrowers is at risk of default. Such a low percentage is quite reasonable given that Grameen Bank is a successful MFI that has been maintaining a high loan payment probability for over three decades. The estimated critical fraction, $t_h=0.0695$, is also small, pointing to two plausible interpretations. First, aside from a very general rationale \cite{Du92}, we had no evidence for letting the number of neighbors saturate at around 100, and therefore may have chosen a level that is too high to reflect reality. If so, the estimate of the threshold $t_h$ should be rather low. Second, because the estimated critical fraction of the average number of neighbors (i.e. 7\% of 100) is about the same as the size of borrowing groups (6 people), it would seem that financial decisions of borrowers are determined by the status of their most immediate neighbors. For a borrower in a critically inactive neighborhood, the probability of extrinsic failure, $p_{ext}=0.0019$ per day, suggests that there is a less than 5.8\% chance for a strategic default to occur in the next month. If the borrower defaults, the average time to recovery consistent with the data is $\overline{\tau}= 39.85$ days. The obtained parameter values put Grameen Bank rather firmly into the domain of regime I (Fig.~\ref{fig05}c), as could be expected from the successful operations of this MFI. However, the low value of $t_h$ is a reason for some concern should the economic environment deteriorate.}

\section*{Discussion}
{\color{black}
Reducing the major credit risks of MFIs to just a few parameters, as we have done herein, offers practical pathways to risk mapping and management, similar in spirit to the approach used by Ref.~\cite{BRB11}. To illustrate this idea, consider that the parameters $t_h$ and $p_{ext}$ describe the risk arising from the latent moral hazard of a strategic default \cite{Mo00, Ma03, CCW07}. In particular, $t_h$ is the critical point at which borrowers start doubting the positive value of further cooperation with the lending program. The value of cooperation may be perceived differently by different individual borrowers, but the important question is how MFIs can strengthen that perception in a general manner. Here we refer to new ideas, independent of the established mechanisms of strict group liability with peer selection and monitoring. The latter mechanisms certainly play an effective role in reducing information asymmetries, but they also transfer the costs of screening and identifying delinquents from lenders to borrowers \cite{St90, Gh99} without guaranteeing lower interest rates in return (especially in the case of a for-profit lender). By contrast, improving the perception of the value of cooperation tends to raise social capital by lowering the transaction costs of working towards common goals \cite{Pr03, Os09}. A good example might be educational activities, which, although primarily aimed at raising human capital, also have the capacity to ``engender goodwill and sentiments of reciprocity'' \cite{KV11}. Furthermore, emphasis on building social capital may have an additional positive effect of lowering the probability of extrinsic failure, $p_{ext}$. This presumed effect is because social capital is based on relations of trust, reciprocity, and connectedness in networks \cite{Pr03}, all of which help not only to establish but also maintain spontaneous cooperation \cite{Os09}.}

{\color{black}
In our numerical simulations, we gave priority to conceptual simplicity over maximum realism. This simplification by no means precludes extending the model within the network theory framework to accommodate additional aspects of reality relevant to microfinance credit risk management. For instance, decaying networks could be used to model permanent defaults \cite{Ma03} by excluding intrinsically failed agents with infinite recovery times. Dynamical networks could also be used to imitate mechanisms of moral hazard complementary to the one we have already described. An example would be failed connections between agents \cite{PMC14} when the information exchange is interrupted as a consequence of lax peer monitoring \cite{Ma03, Co99}. Additionally, new agents could be allowed to enter the network, leading to a class of models that explicitly incorporates outreach as a vital component of microfinance industry \cite{HL11, Co99}. Finally, competition between two or more MFIs \cite{McW05} could be modeled in a setting where networks of borrowers overlap, allowing more robust networks to take over parts of more fragile ones. We therefore conclude that the concept of dynamical networks makes a promising new tool for analyzing open questions in microfinance with the potential for greatly improving, if not revolutionizing, the perception of sustainability in this industry.}

\section*{Methods}
{\color{black}
We constructed an agent-based model mimicking the characteristics of an MFI by surmising that network theory is a natural framework for such a construct. Agents (i.e. equal-sized nodes of the network) represent borrowers, who can be in an active or an inactive state. The active state signifies that the borrower is regularly paying installments, whereas the inactive state indicates that the borrower is currently unable to pay or refusing to continue with the payments. The overall success of the MFI is indicated by the fraction of active agents, $f$, which serves as a proxy for the loan payment probability. An agent can fail intrinsically (i.e. switch from an active to an inactive state) as the involuntary consequence of owning an unsuccessful small enterprise. The probability of intrinsic failure, $p_{int}$, is small because at any moment most enterprises are expected to continue doing business as usual. However, $p_{int}$ increases in a deteriorating economic environment or if the lender is unable to discriminate against risky borrowers with limited collateral (adverse selection). The described formalism is similar to that in Ref.~\cite{MPB14} with the exception that here $p_{int}$ is conditional on agents being in the active state. Otherwise, agents in the inactive state could fail again, which would correspond to multiple defaults of a single borrower. We assume that such multiple defaults are impossible. Finally, because the probability of intrinsic failure is beyond the control of MFIs, this variable can be considered as a forcing for the model.}

{\color{black}
In contrast to intrinsic failure, an owner of a successful small enterprise may be hesitant to pay scheduled installments for extrinsically motivated considerations. These considerations are based on the availability of information on how regularly other borrowers service their loan obligations. We assume that two agents are connected in the usual sense of network theory if each of them knows whether the other one is currently active. Given the way microfinance operations are set up, it is quite certain that such knowledge is common among the members of the same group. The situation becomes progressively more opaque as the geographical reach extends from the group to a management unit, a branch, and, ultimately, the whole MFI. Thus, we set up a hierarchical network in a way that $n_1 < n_2 < n_3 < n_4$ agents corresponding to a group, a unit, a branch, and the whole institution, respectively, are connected with the probabilities $q_1 > q_2 > q_3 > q_4$, respectively. The size parameters ($n_i$) can be constrained by observing Grameen's operations with roughly 6 persons per group, 10 groups per unit, and 7 units per branch. The connectedness probabilities ($q_i$) are more loosely constrained by the facts that the group structure and weekly unit meetings guarantee information exchange, a branch covers a limited geographic area, and the average total number of connections per agent should saturate at around 100~\cite{Du92}. We assume that, as suggested by the Watts model~\cite{Wa02} and additionally explored by Refs.~\cite{MPB14, PMC14}, if a certain critical fraction of neighbors, $t_h$, turns inactive, an agent is tempted---with the small probability $p_{ext}$---to purposely do the same. Such an assumption accounts for the latent moral hazard of strategic default when borrowers no longer believe that other borrowers will continue honoring their loan agreements \cite{Mo00, Ma03, CCW07}. Note that the larger the value of $t_h$, and the smaller the value of $p_{ext}$, the less an agent depends on its neighbors. The parameter $p_{ext}$ is, similarly to the forcing $p_{int}$, conditional on agents being in the active state. Furthermore, the failure is said to be extrinsic because the available information on the state of the other agents prompts a voluntary reaction.}

{\color{black}
In microfinance every failure is addressed by a branch manager. Unsuccessful borrowers may have their payments rescheduled or may be issued new loans to jump-start their businesses. Accordingly, we assume that the inactive agents in the network, irrespective of whether they failed intrinsically or extrinsically, recover after a time $\tau$. Because the manager cannot resolve issues immediately, the time to recovery is greater than a certain minimum time, $\tau_0$. Managers may also vary in their ability to handle the problems that led to the failure, which is why the recovery time has a random component~\cite{PMC14}, $\Delta \tau$, so that $\tau = \tau_0 + \Delta \tau$. The random variable $\Delta \tau$ is assumed to be exponentially distributed with the standard deviation $\sigma$, a measure of the diverse abilities of the managers. One of the properties of the exponential distribution is that $\sigma$ also serves as the expectation of the random variable $\Delta \tau$, thus giving the expected time to recovery $\overline{\tau} = \tau_0 + \sigma$. With these specifications, we are finally in a position to define that an agent is in the inactive state if it has failed either intrinsically or extrinsically, and if the time since failure is less than the time to recovery. Otherwise, the agent is in the active state. A reader keen on better understanding the intricacies of the described model may benefit from the derivation of a deterministic analogue below. Mathematical symbols appearing throughout are given in Table~\ref{tab01}.} 

\subsection*{Deterministic model}
{\color{black}
We derive deterministic equations that govern the time evolution of the average number of active, intrinsically failed, and extrinsically failed agents. Let us denote these quantities by $f(t)$, $n_{int}(t)$, and $n_{ext}(t)$, respectively. The change in $f(t)$ is given by the difference in the total number of recovering, $R_{tot}(t)$, and failing, $F_{tot}(t)$, agents at an arbitrary moment $t$. For recoveries, we have $R_{tot}(t) = [1-f(t-\overline{\tau})] / \overline{\tau}$ because out of $1-f(t-\overline{\tau})$ agents that are in the failed state at $t-\overline{\tau}$, only the fraction $1/\overline{\tau}$ can recover at $t$. For failures, we have $F_{tot}(t) = F_{int}(t) + F_{ext}(t) = f(t) p_{int} + f(t) (1-p_{int}) E(t) p_{ext}$ because (i) only active agents can fail, (ii) intrinsic failures have priority over extrinsic ones (i.e. the latter are conditional on the former not happening at the same time), and (iii) the probability of an agent having a critically inactive neighborhood is $E(t)$. Furthermore, it holds by definition that $R_{tot}(t+\overline{\tau}) = F_{tot}(t)$ which yields
\begin{equation}
 f(t) = \frac{1}{1 + [p_{int} + E(t) (p_{ext} - p_{int} p_{ext})] \overline{\tau}}.
\label{eqS01}
\end{equation}
It is now apparent that the equilibrium fraction of active agents, $f^*$---given in Eq.~(\ref{eq01})---follows from Eq.~(\ref{eqS01}) as the time dependence of the probability $E(t)$ wanes. Next, looking at the number of recoveries from intrinsic, $R_{int}(t)$, and extrinsic, $R_{ext}(t)$, failures separately, the same logic as before dictates $R_{int}(t) = n_{int}(t-\overline{\tau}) / \overline{\tau}$ and $R_{ext}(t) = n_{ext}(t-\overline{\tau}) / \overline{\tau}$. The definitions imply $R_{int}(t+\overline{\tau}) = F_{int}(t)$ and $R_{ext}(t+\overline{\tau}) = F_{ext}(t)$, resulting in
\begin{subequations}
\begin{align}
 n_{int}(t) &= \overline{\tau} f(t) p_{int},                    \label{eqS02a} \\
 n_{ext}(t) &= \overline{\tau} f(t) (1 - p_{int}) E(t) p_{ext}. \label{eqS02b}
\end{align}
\end{subequations}
Using Eqs.~(\ref{eqS01}, \ref{eqS02a}, and \ref{eqS02b}) a quick check gives $f(t)+n_{int}(t)+n_{ext}(t) = 1$ as it should be. We thus see that the underlying dynamics of our model are relatively simple, yet able to generate rather complex dynamical phenomena owing to the stochasticity and a small set of rules that determine how $E(t)$ depends on time.}

\subsection*{Empirical analysis}
{\color{black}
Real-world microfinance data can be interpreted in the context of our modeling framework by estimating the model parameters from the data. An appropriate method for doing so is the method of moments. We apply this method to a 10-year monthly data set, consisting of loan collection probabilities recorded by Grameen Bank from June 2002 to January 2013. Stated more precisely, ``loan collection probabilities'' mean the percentage of the amounts due that were actually paid, which is compatible with the model output $f$. One difficulty in estimating the model parameters is the apparent trend in the data (Fig.~\ref{fig05}a), suggesting a time-dependent forcing of the form $p_{int}(t) = \overline{p}_{int} + \Delta p_{int}(t)$, where $\overline{p}_{int}$ is a constant and $\Delta p_{int}(t)$ a zero-mean smooth function of time. We therefore use a smoothing spline, $s(t)$, to extract the trend and then, during the estimation, treat $\overline{p}_{int}$ as just another parameter. To keep the total number of parameters at a computationally manageable level, we assume without a major loss of generality that the minimum time to recovery, $\tau_0$, is seven days. Altogether we aim at reproducing the first four moments of the available data set, i.e. the mean ($m_1$), the standard deviation ($m_2^{1/2}$), the standardized skewness ($m_3/m_2^{3/2}$), and the standardized kurtosis ($m_4/m_2^{2}$). At the point of best fit, the empirical moments compare favorably with the model-generated ones (mean $\pm$ s.d. from 500 runs): 0.9805 vs. $0.9833 \pm 0.0004$, 0.0023 vs. $0.0023 \pm 0.0002$, -0.9133 vs. $-0.2907 \pm 0.2705$, and 4.0024 vs. $3.0238 \pm 0.5994$ for the mean, the standard deviation, the standardized skewness, and the standardized kurtosis, respectively. A comparison of the original data set with three typical model runs is shown in Fig.~\ref{fig05}b. In these simulations, we assume that the time-dependent forcing is $\Delta p_{int}(t) = c [s(t) - \overline{s}]$, where $c$ is a proportionality constant set to $-\frac{1}{8}$ and $\overline{s}$ is the average of $s(t)$. Such an assumption is justifiable in regime I (Fig.~\ref{fig01}). Finally, with reasonable agreement between the data and the model results, we construct a risk map for Grameen bank analogous to those shown in Fig.~\ref{fig03}. In this case, however, we use a $p_{ext}$-$p_{int}$ phase diagram (Fig.~\ref{fig05}c), where the probability of intrinsic failure should be interpreted as the average forcing, $\overline{p}_{int}$.}

\section*{Acknowledgments}
We are grateful to Ashir Ahmed, Sa\v{s}a Drezgi\'{c}, Simon Levin, Tadasu Matsuo, and Zhen Wang for valuable discussions. We acknowledge support from (i) Kyushu University Graduate Education and Research Training Program in Decision Science for a Sustainable Society no. P02, (ii) Japan Society for the Promotion of Science (JSPS) Postdoctoral Fellowship Program for Foreign Researchers no. P13380 and an accompanying Grant-in-Aid for Scientific Research, (iii) University of Rijeka project no. 13.05.1.3.05, and (iv) JSPS Grant-in-Aid for General Scientific Research B no. 24370011.

\bibliography{Literature}

\begin{thebibliography}{35}%
\makeatletter
\providecommand \@ifxundefined [1]{%
 \@ifx{#1\undefined}
}%
\providecommand \@ifnum [1]{%
 \ifnum #1\expandafter \@firstoftwo
 \else \expandafter \@secondoftwo
 \fi
}%
\providecommand \@ifx [1]{%
 \ifx #1\expandafter \@firstoftwo
 \else \expandafter \@secondoftwo
 \fi
}%
\providecommand \natexlab [1]{#1}%
\providecommand \enquote  [1]{``#1''}%
\providecommand \bibnamefont  [1]{#1}%
\providecommand \bibfnamefont [1]{#1}%
\providecommand \citenamefont [1]{#1}%
\providecommand \href@noop [0]{\@secondoftwo}%
\providecommand \href [0]{\begingroup \@sanitize@url \@href}%
\providecommand \@href[1]{\@@startlink{#1}\@@href}%
\providecommand \@@href[1]{\endgroup#1\@@endlink}%
\providecommand \@sanitize@url [0]{\catcode `\\12\catcode `\$12\catcode
  `\&12\catcode `\#12\catcode `\^12\catcode `\_12\catcode `\%12\relax}%
\providecommand \@@startlink[1]{}%
\providecommand \@@endlink[0]{}%
\providecommand \url  [0]{\begingroup\@sanitize@url \@url }%
\providecommand \@url [1]{\endgroup\@href {#1}{\urlprefix }}%
\providecommand \urlprefix  [0]{URL }%
\providecommand \Eprint [0]{\href }%
\providecommand \doibase [0]{http://dx.doi.org/}%
\providecommand \selectlanguage [0]{\@gobble}%
\providecommand \bibinfo  [0]{\@secondoftwo}%
\providecommand \bibfield  [0]{\@secondoftwo}%
\providecommand \translation [1]{[#1]}%
\providecommand \BibitemOpen [0]{}%
\providecommand \bibitemStop [0]{}%
\providecommand \bibitemNoStop [0]{.\EOS\space}%
\providecommand \EOS [0]{\spacefactor3000\relax}%
\providecommand \BibitemShut  [1]{\csname bibitem#1\endcsname}%
\let\auto@bib@innerbib\@empty
\bibitem [{\citenamefont {Morduch}(199)}]{Mo99}%
  \BibitemOpen
  \bibfield  {author} {\bibinfo {author} {\bibfnamefont {J.}~\bibnamefont
  {Morduch}},\ }\href@noop {} {\bibfield  {journal} {\bibinfo  {journal} {J
  Econ Lit}\ }\textbf {\bibinfo {volume} {37}},\ \bibinfo {pages} {1569}
  (\bibinfo {year} {199})}\BibitemShut {NoStop}%
\bibitem [{\citenamefont {Copestake}\ \emph {et~al.}(2001)\citenamefont
  {Copestake}, \citenamefont {Bhalotra},\ and\ \citenamefont
  {Johnson}}]{CBJ01}%
  \BibitemOpen
  \bibfield  {author} {\bibinfo {author} {\bibfnamefont {J.}~\bibnamefont
  {Copestake}}, \bibinfo {author} {\bibfnamefont {S.}~\bibnamefont {Bhalotra}},
  \ and\ \bibinfo {author} {\bibfnamefont {S.}~\bibnamefont {Johnson}},\
  }\href@noop {} {\bibfield  {journal} {\bibinfo  {journal} {J Dev Stud}\
  }\textbf {\bibinfo {volume} {37}},\ \bibinfo {pages} {81} (\bibinfo {year}
  {2001})}\BibitemShut {NoStop}%
\bibitem [{\citenamefont {Khandker}(2005)}]{Kh05}%
  \BibitemOpen
  \bibfield  {author} {\bibinfo {author} {\bibfnamefont {S.~R.}\ \bibnamefont
  {Khandker}},\ }\href@noop {} {\bibfield  {journal} {\bibinfo  {journal}
  {World Bank Econ Rev}\ }\textbf {\bibinfo {volume} {19}},\ \bibinfo {pages}
  {263} (\bibinfo {year} {2005})}\BibitemShut {NoStop}%
\bibitem [{\citenamefont {Hermes}\ and\ \citenamefont {Lensink}(2011)}]{HL11}%
  \BibitemOpen
  \bibfield  {author} {\bibinfo {author} {\bibfnamefont {N.}~\bibnamefont
  {Hermes}}\ and\ \bibinfo {author} {\bibfnamefont {R.}~\bibnamefont
  {Lensink}},\ }\href@noop {} {\bibfield  {journal} {\bibinfo  {journal} {World
  Dev}\ }\textbf {\bibinfo {volume} {39}},\ \bibinfo {pages} {875} (\bibinfo
  {year} {2011})}\BibitemShut {NoStop}%
\bibitem [{\citenamefont {Karlan}\ and\ \citenamefont {Zinman}(2010)}]{KZ10}%
  \BibitemOpen
  \bibfield  {author} {\bibinfo {author} {\bibfnamefont {D.}~\bibnamefont
  {Karlan}}\ and\ \bibinfo {author} {\bibfnamefont {J.}~\bibnamefont
  {Zinman}},\ }\href@noop {} {\bibfield  {journal} {\bibinfo  {journal} {Rev
  Financ Stud}\ }\textbf {\bibinfo {volume} {23}},\ \bibinfo {pages} {433}
  (\bibinfo {year} {2010})}\BibitemShut {NoStop}%
\bibitem [{\citenamefont {Marr}(2003)}]{Ma03}%
  \BibitemOpen
  \bibfield  {author} {\bibinfo {author} {\bibfnamefont {A.}~\bibnamefont
  {Marr}},\ }\href@noop {} {\bibfield  {journal} {\bibinfo  {journal} {J
  Microfinance}\ }\textbf {\bibinfo {volume} {5}},\ \bibinfo {pages} {7}
  (\bibinfo {year} {2003})}\BibitemShut {NoStop}%
\bibitem [{\citenamefont {Elahi}\ and\ \citenamefont
  {Danopoulos}(2004)}]{ED04}%
  \BibitemOpen
  \bibfield  {author} {\bibinfo {author} {\bibfnamefont {K.~Q.}\ \bibnamefont
  {Elahi}}\ and\ \bibinfo {author} {\bibfnamefont {C.~P.}\ \bibnamefont
  {Danopoulos}},\ }\href@noop {} {\bibfield  {journal} {\bibinfo  {journal} {J
  Polit Mil Soc}\ }\textbf {\bibinfo {volume} {32}},\ \bibinfo {pages} {61}
  (\bibinfo {year} {2004})}\BibitemShut {NoStop}%
\bibitem [{\citenamefont {Hermes}\ and\ \citenamefont {Lensink}(2007)}]{HL07}%
  \BibitemOpen
  \bibfield  {author} {\bibinfo {author} {\bibfnamefont {N.}~\bibnamefont
  {Hermes}}\ and\ \bibinfo {author} {\bibfnamefont {R.}~\bibnamefont
  {Lensink}},\ }\href@noop {} {\bibfield  {journal} {\bibinfo  {journal} {Econ
  J}\ }\textbf {\bibinfo {volume} {117}},\ \bibinfo {pages} {F1} (\bibinfo
  {year} {2007})}\BibitemShut {NoStop}%
\bibitem [{\citenamefont {Conning}(1999)}]{Co99}%
  \BibitemOpen
  \bibfield  {author} {\bibinfo {author} {\bibfnamefont {J.}~\bibnamefont
  {Conning}},\ }\href@noop {} {\bibfield  {journal} {\bibinfo  {journal} {J Dev
  Econ}\ }\textbf {\bibinfo {volume} {60}},\ \bibinfo {pages} {51} (\bibinfo
  {year} {1999})}\BibitemShut {NoStop}%
\bibitem [{\citenamefont {McIntosh}\ and\ \citenamefont
  {Wydick}(2005)}]{McW05}%
  \BibitemOpen
  \bibfield  {author} {\bibinfo {author} {\bibfnamefont {C.}~\bibnamefont
  {McIntosh}}\ and\ \bibinfo {author} {\bibfnamefont {B.}~\bibnamefont
  {Wydick}},\ }\href@noop {} {\bibfield  {journal} {\bibinfo  {journal} {J Dev
  Econ}\ }\textbf {\bibinfo {volume} {78}},\ \bibinfo {pages} {271} (\bibinfo
  {year} {2005})}\BibitemShut {NoStop}%
\bibitem [{\citenamefont {Mersland}\ and\ \citenamefont
  {Str{\o}m}(2010)}]{MS10}%
  \BibitemOpen
  \bibfield  {author} {\bibinfo {author} {\bibfnamefont {R.}~\bibnamefont
  {Mersland}}\ and\ \bibinfo {author} {\bibfnamefont {R.~{\O}.}\ \bibnamefont
  {Str{\o}m}},\ }\href@noop {} {\bibfield  {journal} {\bibinfo  {journal}
  {World Dev}\ }\textbf {\bibinfo {volume} {38}},\ \bibinfo {pages} {28}
  (\bibinfo {year} {2010})}\BibitemShut {NoStop}%
\bibitem [{\citenamefont {Dehejia}\ \emph {et~al.}(2012)\citenamefont
  {Dehejia}, \citenamefont {Montgomery},\ and\ \citenamefont
  {Morduch}}]{DMM05}%
  \BibitemOpen
  \bibfield  {author} {\bibinfo {author} {\bibfnamefont {R.}~\bibnamefont
  {Dehejia}}, \bibinfo {author} {\bibfnamefont {H.}~\bibnamefont {Montgomery}},
  \ and\ \bibinfo {author} {\bibfnamefont {J.}~\bibnamefont {Morduch}},\
  }\href@noop {} {\bibfield  {journal} {\bibinfo  {journal} {J Dev Econ}\
  }\textbf {\bibinfo {volume} {97}},\ \bibinfo {pages} {437} (\bibinfo {year}
  {2012})}\BibitemShut {NoStop}%
\bibitem [{\citenamefont {May}\ \emph {et~al.}(2008)\citenamefont {May},
  \citenamefont {Levin},\ and\ \citenamefont {Sugihara}}]{MLS08}%
  \BibitemOpen
  \bibfield  {author} {\bibinfo {author} {\bibfnamefont {R.~M.}\ \bibnamefont
  {May}}, \bibinfo {author} {\bibfnamefont {S.~A.}\ \bibnamefont {Levin}}, \
  and\ \bibinfo {author} {\bibfnamefont {G.}~\bibnamefont {Sugihara}},\
  }\href@noop {} {\bibfield  {journal} {\bibinfo  {journal} {Nature}\ }\textbf
  {\bibinfo {volume} {451}},\ \bibinfo {pages} {893} (\bibinfo {year}
  {2008})}\BibitemShut {NoStop}%
\bibitem [{\citenamefont {May}\ and\ \citenamefont
  {Arinaminpathy}(2010)}]{MA10}%
  \BibitemOpen
  \bibfield  {author} {\bibinfo {author} {\bibfnamefont {R.~M.}\ \bibnamefont
  {May}}\ and\ \bibinfo {author} {\bibfnamefont {N.}~\bibnamefont
  {Arinaminpathy}},\ }\href@noop {} {\bibfield  {journal} {\bibinfo  {journal}
  {J R Soc Interface}\ }\textbf {\bibinfo {volume} {7}},\ \bibinfo {pages}
  {823} (\bibinfo {year} {2010})}\BibitemShut {NoStop}%
\bibitem [{\citenamefont {Beale}\ \emph {et~al.}(2011)\citenamefont {Beale},
  \citenamefont {Rand}, \citenamefont {Battey}, \citenamefont {Croxson},
  \citenamefont {May},\ and\ \citenamefont {Nowak}}]{BRB11}%
  \BibitemOpen
  \bibfield  {author} {\bibinfo {author} {\bibfnamefont {N.}~\bibnamefont
  {Beale}}, \bibinfo {author} {\bibfnamefont {D.~G.}\ \bibnamefont {Rand}},
  \bibinfo {author} {\bibfnamefont {H.}~\bibnamefont {Battey}}, \bibinfo
  {author} {\bibfnamefont {K.}~\bibnamefont {Croxson}}, \bibinfo {author}
  {\bibfnamefont {R.~M.}\ \bibnamefont {May}}, \ and\ \bibinfo {author}
  {\bibfnamefont {M.~A.}\ \bibnamefont {Nowak}},\ }\href@noop {} {\bibfield
  {journal} {\bibinfo  {journal} {Proc. Natl. Acad. Sci. USA}\ }\textbf
  {\bibinfo {volume} {108}},\ \bibinfo {pages} {12647} (\bibinfo {year}
  {2011})}\BibitemShut {NoStop}%
\bibitem [{\citenamefont {Haldane}\ and\ \citenamefont {May}(2011)}]{HM11}%
  \BibitemOpen
  \bibfield  {author} {\bibinfo {author} {\bibfnamefont {A.~G.}\ \bibnamefont
  {Haldane}}\ and\ \bibinfo {author} {\bibfnamefont {R.~M.}\ \bibnamefont
  {May}},\ }\href@noop {} {\bibfield  {journal} {\bibinfo  {journal} {Nature}\
  }\textbf {\bibinfo {volume} {469}},\ \bibinfo {pages} {351} (\bibinfo {year}
  {2011})}\BibitemShut {NoStop}%
\bibitem [{\citenamefont {Arinaminpathy}\ \emph {et~al.}(2012)\citenamefont
  {Arinaminpathy}, \citenamefont {Kapadia},\ and\ \citenamefont {May}}]{AKM12}%
  \BibitemOpen
  \bibfield  {author} {\bibinfo {author} {\bibfnamefont {N.}~\bibnamefont
  {Arinaminpathy}}, \bibinfo {author} {\bibfnamefont {S.}~\bibnamefont
  {Kapadia}}, \ and\ \bibinfo {author} {\bibfnamefont {R.~M.}\ \bibnamefont
  {May}},\ }\href@noop {} {\bibfield  {journal} {\bibinfo  {journal} {Proc.
  Natl. Acad. Sci. USA}\ }\textbf {\bibinfo {volume} {109}},\ \bibinfo {pages}
  {18338} (\bibinfo {year} {2012})}\BibitemShut {NoStop}%
\bibitem [{\citenamefont {Gale}\ and\ \citenamefont {Kariv}(2007)}]{GK07}%
  \BibitemOpen
  \bibfield  {author} {\bibinfo {author} {\bibfnamefont {D.~M.}\ \bibnamefont
  {Gale}}\ and\ \bibinfo {author} {\bibfnamefont {S.}~\bibnamefont {Kariv}},\
  }\href@noop {} {\bibfield  {journal} {\bibinfo  {journal} {Am Econ Rev}\
  }\textbf {\bibinfo {volume} {97}},\ \bibinfo {pages} {99} (\bibinfo {year}
  {2007})}\BibitemShut {NoStop}%
\bibitem [{\citenamefont {Schweitzer}\ \emph {et~al.}(2009)\citenamefont
  {Schweitzer}, \citenamefont {Fagiolo}, \citenamefont {Sornette},
  \citenamefont {Vega-Redondo}, \citenamefont {Vespignani},\ and\ \citenamefont
  {White}}]{SFS09}%
  \BibitemOpen
  \bibfield  {author} {\bibinfo {author} {\bibfnamefont {F.}~\bibnamefont
  {Schweitzer}}, \bibinfo {author} {\bibfnamefont {G.}~\bibnamefont {Fagiolo}},
  \bibinfo {author} {\bibfnamefont {D.}~\bibnamefont {Sornette}}, \bibinfo
  {author} {\bibfnamefont {F.}~\bibnamefont {Vega-Redondo}}, \bibinfo {author}
  {\bibfnamefont {A.}~\bibnamefont {Vespignani}}, \ and\ \bibinfo {author}
  {\bibfnamefont {D.~R.}\ \bibnamefont {White}},\ }\href@noop {} {\bibfield
  {journal} {\bibinfo  {journal} {Science}\ }\textbf {\bibinfo {volume}
  {325}},\ \bibinfo {pages} {422} (\bibinfo {year} {2009})}\BibitemShut
  {NoStop}%
\bibitem [{\citenamefont {Gai}\ and\ \citenamefont {Kapadia}(2010)}]{GK10}%
  \BibitemOpen
  \bibfield  {author} {\bibinfo {author} {\bibfnamefont {P.}~\bibnamefont
  {Gai}}\ and\ \bibinfo {author} {\bibfnamefont {S.}~\bibnamefont {Kapadia}},\
  }\href@noop {} {\bibfield  {journal} {\bibinfo  {journal} {Proc R Soc A}\
  }\textbf {\bibinfo {volume} {466}},\ \bibinfo {pages} {2401} (\bibinfo {year}
  {2010})}\BibitemShut {NoStop}%
\bibitem [{\citenamefont {Elliott}\ \emph {et~al.}(2014)\citenamefont
  {Elliott}, \citenamefont {Golub},\ and\ \citenamefont {Jackson}}]{EGJ14}%
  \BibitemOpen
  \bibfield  {author} {\bibinfo {author} {\bibfnamefont {M.}~\bibnamefont
  {Elliott}}, \bibinfo {author} {\bibfnamefont {B.}~\bibnamefont {Golub}}, \
  and\ \bibinfo {author} {\bibfnamefont {M.~O.}\ \bibnamefont {Jackson}},\
  }\href@noop {} {\bibfield  {journal} {\bibinfo  {journal} {Am Econ Rev}\
  }\textbf {\bibinfo {volume} {104}},\ \bibinfo {pages} {3115} (\bibinfo {year}
  {2014})}\BibitemShut {NoStop}%
\bibitem [{\citenamefont {Majdandzic}\ \emph {et~al.}(2014)\citenamefont
  {Majdandzic}, \citenamefont {Podobnik}, \citenamefont {Buldyrev},
  \citenamefont {Kenett}, \citenamefont {Havlin},\ and\ \citenamefont
  {Stanley}}]{MPB14}%
  \BibitemOpen
  \bibfield  {author} {\bibinfo {author} {\bibfnamefont {A.}~\bibnamefont
  {Majdandzic}}, \bibinfo {author} {\bibfnamefont {B.}~\bibnamefont
  {Podobnik}}, \bibinfo {author} {\bibfnamefont {S.~V.}\ \bibnamefont
  {Buldyrev}}, \bibinfo {author} {\bibfnamefont {D.~Y.}\ \bibnamefont
  {Kenett}}, \bibinfo {author} {\bibfnamefont {S.}~\bibnamefont {Havlin}}, \
  and\ \bibinfo {author} {\bibfnamefont {H.~E.}\ \bibnamefont {Stanley}},\
  }\href@noop {} {\bibfield  {journal} {\bibinfo  {journal} {Nature Phys}\
  }\textbf {\bibinfo {volume} {10}},\ \bibinfo {pages} {34} (\bibinfo {year}
  {2014})}\BibitemShut {NoStop}%
\bibitem [{\citenamefont {Podobnik}\ \emph
  {et~al.}(2014{\natexlab{a}})\citenamefont {Podobnik}, \citenamefont
  {Majdandzic}, \citenamefont {Curme}, \citenamefont {Qiao}, \citenamefont
  {Zhou}, \citenamefont {Stanley},\ and\ \citenamefont {Li}}]{PMC14}%
  \BibitemOpen
  \bibfield  {author} {\bibinfo {author} {\bibfnamefont {B.}~\bibnamefont
  {Podobnik}}, \bibinfo {author} {\bibfnamefont {A.}~\bibnamefont
  {Majdandzic}}, \bibinfo {author} {\bibfnamefont {C.}~\bibnamefont {Curme}},
  \bibinfo {author} {\bibfnamefont {Z.}~\bibnamefont {Qiao}}, \bibinfo {author}
  {\bibfnamefont {W.~X.}\ \bibnamefont {Zhou}}, \bibinfo {author}
  {\bibfnamefont {H.~E.}\ \bibnamefont {Stanley}}, \ and\ \bibinfo {author}
  {\bibfnamefont {B.}~\bibnamefont {Li}},\ }\href@noop {} {\bibfield  {journal}
  {\bibinfo  {journal} {Phys Rev E}\ }\textbf {\bibinfo {volume} {89}},\
  \bibinfo {pages} {042807} (\bibinfo {year} {2014}{\natexlab{a}})}\BibitemShut
  {NoStop}%
\bibitem [{\citenamefont {Podobnik}\ \emph
  {et~al.}(2014{\natexlab{b}})\citenamefont {Podobnik}, \citenamefont
  {Horvatic}, \citenamefont {Ling}, \citenamefont {Bertella},\ and\
  \citenamefont {Li}}]{PHL14}%
  \BibitemOpen
  \bibfield  {author} {\bibinfo {author} {\bibfnamefont {B.}~\bibnamefont
  {Podobnik}}, \bibinfo {author} {\bibfnamefont {D.}~\bibnamefont {Horvatic}},
  \bibinfo {author} {\bibfnamefont {F.}~\bibnamefont {Ling}}, \bibinfo {author}
  {\bibfnamefont {M.}~\bibnamefont {Bertella}}, \ and\ \bibinfo {author}
  {\bibfnamefont {B.}~\bibnamefont {Li}},\ }\href@noop {} {\bibfield  {journal}
  {\bibinfo  {journal} {Europhys Lett}\ }\textbf {\bibinfo {volume} {106}},\
  \bibinfo {pages} {68003} (\bibinfo {year} {2014}{\natexlab{b}})}\BibitemShut
  {NoStop}%
\bibitem [{\citenamefont {Stiglitz}(1990)}]{St90}%
  \BibitemOpen
  \bibfield  {author} {\bibinfo {author} {\bibfnamefont {J.~E.}\ \bibnamefont
  {Stiglitz}},\ }\href@noop {} {\bibfield  {journal} {\bibinfo  {journal}
  {World Bank Econ Rev}\ }\textbf {\bibinfo {volume} {4}},\ \bibinfo {pages}
  {351} (\bibinfo {year} {1990})}\BibitemShut {NoStop}%
\bibitem [{\citenamefont {Watts}(2002)}]{Wa02}%
  \BibitemOpen
  \bibfield  {author} {\bibinfo {author} {\bibfnamefont {D.~J.}\ \bibnamefont
  {Watts}},\ }\href@noop {} {\bibfield  {journal} {\bibinfo  {journal} {Proc
  Natl Acad Sci USA}\ }\textbf {\bibinfo {volume} {99}},\ \bibinfo {pages}
  {5766} (\bibinfo {year} {2002})}\BibitemShut {NoStop}%
\bibitem [{\citenamefont {Morduch}(2000)}]{Mo00}%
  \BibitemOpen
  \bibfield  {author} {\bibinfo {author} {\bibfnamefont {J.}~\bibnamefont
  {Morduch}},\ }\href@noop {} {\bibfield  {journal} {\bibinfo  {journal} {World
  Dev}\ }\textbf {\bibinfo {volume} {28}},\ \bibinfo {pages} {617} (\bibinfo
  {year} {2000})}\BibitemShut {NoStop}%
\bibitem [{\citenamefont {Cassar}\ \emph {et~al.}(2007)\citenamefont {Cassar},
  \citenamefont {Crowley},\ and\ \citenamefont {Wydick}}]{CCW07}%
  \BibitemOpen
  \bibfield  {author} {\bibinfo {author} {\bibfnamefont {A.}~\bibnamefont
  {Cassar}}, \bibinfo {author} {\bibfnamefont {L.}~\bibnamefont {Crowley}}, \
  and\ \bibinfo {author} {\bibfnamefont {B.}~\bibnamefont {Wydick}},\
  }\href@noop {} {\bibfield  {journal} {\bibinfo  {journal} {Econ J}\ }\textbf
  {\bibinfo {volume} {117}},\ \bibinfo {pages} {F85} (\bibinfo {year}
  {2007})}\BibitemShut {NoStop}%
\bibitem [{\citenamefont {Dunbar}(1992)}]{Du92}%
  \BibitemOpen
  \bibfield  {author} {\bibinfo {author} {\bibfnamefont {R.~I.}\ \bibnamefont
  {Dunbar}},\ }\href@noop {} {\bibfield  {journal} {\bibinfo  {journal} {J Hum
  Evol}\ }\textbf {\bibinfo {volume} {22}},\ \bibinfo {pages} {469} (\bibinfo
  {year} {1992})}\BibitemShut {NoStop}%
\bibitem [{\citenamefont {Scheffer}(2010)}]{Sc10}%
  \BibitemOpen
  \bibfield  {author} {\bibinfo {author} {\bibfnamefont {M.}~\bibnamefont
  {Scheffer}},\ }\href@noop {} {\bibfield  {journal} {\bibinfo  {journal}
  {Nature}\ }\textbf {\bibinfo {volume} {467}},\ \bibinfo {pages} {411}
  (\bibinfo {year} {2010})}\BibitemShut {NoStop}%
\bibitem [{\citenamefont {Peng}\ \emph {et~al.}(1994)\citenamefont {Peng},
  \citenamefont {Buldyrev}, \citenamefont {Havlin}, \citenamefont {Simons},
  \citenamefont {Stanley},\ and\ \citenamefont {Goldberger}}]{PBH94}%
  \BibitemOpen
  \bibfield  {author} {\bibinfo {author} {\bibfnamefont {C.~K.}\ \bibnamefont
  {Peng}}, \bibinfo {author} {\bibfnamefont {S.~V.}\ \bibnamefont {Buldyrev}},
  \bibinfo {author} {\bibfnamefont {S.}~\bibnamefont {Havlin}}, \bibinfo
  {author} {\bibfnamefont {M.}~\bibnamefont {Simons}}, \bibinfo {author}
  {\bibfnamefont {H.~E.}\ \bibnamefont {Stanley}}, \ and\ \bibinfo {author}
  {\bibfnamefont {A.~L.}\ \bibnamefont {Goldberger}},\ }\href@noop {}
  {\bibfield  {journal} {\bibinfo  {journal} {Phys Rev E}\ }\textbf {\bibinfo
  {volume} {49}},\ \bibinfo {pages} {1685} (\bibinfo {year}
  {1994})}\BibitemShut {NoStop}%
\bibitem [{\citenamefont {Ghatak}(1999)}]{Gh99}%
  \BibitemOpen
  \bibfield  {author} {\bibinfo {author} {\bibfnamefont {M.}~\bibnamefont
  {Ghatak}},\ }\href@noop {} {\bibfield  {journal} {\bibinfo  {journal} {J Dev
  Econ}\ }\textbf {\bibinfo {volume} {60}},\ \bibinfo {pages} {27} (\bibinfo
  {year} {1999})}\BibitemShut {NoStop}%
\bibitem [{\citenamefont {Pretty}(2003)}]{Pr03}%
  \BibitemOpen
  \bibfield  {author} {\bibinfo {author} {\bibfnamefont {J.}~\bibnamefont
  {Pretty}},\ }\href@noop {} {\bibfield  {journal} {\bibinfo  {journal}
  {Science}\ }\textbf {\bibinfo {volume} {302}},\ \bibinfo {pages} {1912}
  (\bibinfo {year} {2003})}\BibitemShut {NoStop}%
\bibitem [{\citenamefont {Ostrom}(2009)}]{Os09}%
  \BibitemOpen
  \bibfield  {author} {\bibinfo {author} {\bibfnamefont {E.}~\bibnamefont
  {Ostrom}},\ }\href@noop {} {\bibfield  {journal} {\bibinfo  {journal}
  {Science}\ }\textbf {\bibinfo {volume} {325}},\ \bibinfo {pages} {419–}
  (\bibinfo {year} {2009})}\BibitemShut {NoStop}%
\bibitem [{\citenamefont {Karlan}\ and\ \citenamefont {Valdivia}(2011)}]{KV11}%
  \BibitemOpen
  \bibfield  {author} {\bibinfo {author} {\bibfnamefont {D.}~\bibnamefont
  {Karlan}}\ and\ \bibinfo {author} {\bibfnamefont {M.}~\bibnamefont
  {Valdivia}},\ }\href@noop {} {\bibfield  {journal} {\bibinfo  {journal} {Rev
  Econ Stat}\ }\textbf {\bibinfo {volume} {93}},\ \bibinfo {pages} {510}
  (\bibinfo {year} {2011})}\BibitemShut {NoStop}%
\end{thebibliography}%

\end{document}